\documentclass[prd,twocolumn,preprintnumbers,superscriptaddress,nofootinbib]{revtex4}
\usepackage[a4paper, hdivide={1.91cm,,1.165cm}, vdivide={1.83cm,,2.6cm}]{geometry}

\usepackage{amstext,amssymb}
\usepackage{amsmath}
\usepackage{graphicx}
\usepackage[hyperfootnotes=false]{hyperref}
\usepackage{xspace}
\usepackage{color}
\usepackage{units}
\usepackage{slashed} 
\usepackage{braket}

\usepackage{wrapfig}
\definecolor{light-gray}{gray}{0.95}
\usepackage{tcolorbox}
\begin{document}

\title{MSW effect with quark matter: Neutron Star as a case study}

\author{Hiranmaya \surname{Mishra}}
\email{hm@prl.res.in}
\affiliation{Physical Research Laboratory, Navarangpura, Ahmedabad, India}
\affiliation{School of Physical Sciences, National Institute of Science Education and Research Bhubanwswar, HBNI, Jatni 752050,Odisha, India}
\author{Prasanta K. \surname{Panigrahi}}
\email{panigrahi.iiser@gmail.com}
\affiliation{Indian Institute of Science Education and Research Kolkata, India}
\author{Sudhanwa \surname{Patra}}
\email{sudhanwa@iitbhilai.ac.in}
\affiliation{Department of Physics, Indian Institute of Technology Bhilai, Raipur, India}
\author{Utpal \surname{Sarkar}}
\email{utpal.sarkar.prl@gmail.com}
\affiliation{Indian Institute of Science Education and Research Kolkata, India}

\begin{abstract}
With the recent findings from various astrophysical results hinting towards possible existence of strange quark matters with the baryonic
resonances such as $\Lambda^0, \Sigma^0, \Xi, \Omega$  in the core of neutron stars, we investigate the MSW effect, in general, in quark matter. We find that the resonance condition for the complete conversion of down-quark to strange quark requires estremely large matter density ($\rho_u \simeq 10^{5}\,\mbox{fm}^{-3} $). Nonetheless the neutron stars provide a best condition for the conversion to be statistically  significant which is of the same order as is expected from imposing charge neutrality condition. This has a possibility of resolving the hyperon puzzle as well as the equation of state for dense baryonic matter.

\vspace*{-1.0cm}
\end{abstract}


\maketitle

\noindent
%
The paramount discovery of neutrino oscillation confirming that the neutrinos have non-zero mass and they do change identities while propagating brought the important difference to view of the universe~\cite{Super-Kamiokande:2005wtt,SNO:2002tuh}
and the physics beyond the Standard Model (SM) of particle physics. With times, Mikheyev, Smirnov and Wolfenstein (MSW)~\cite{Wolfenstein:1977ue,Mikheyev:1985zog,Bethe:1986ej,Smirnov:2004zv} mechanism is found to be very popular due to its potential to explain the flavor conversion of solar neutrinos during their propagation in solar matter even with the small vaccum mixing angle. The basic idea of MSW effect is that neutrino while propagating in matter are subjected to a potential arising from the coherent elastic  scattering with the particles (protons, neutrons and electrons) through charged current interaction. Even with small vaccum mixing angle, the matter potential which acts as an index of refraction modifies the in-medium mixing angle and can play an important role in flavor conversion of neutrinos. It is well know that for solar neutrinos with energy around MeV, the estimated mean free path of the neutrinos in normal matter ($\rho \sim 10^{-15}\, \mbox{fm}^{-3}$) is about  
$10^{14}$\,km which at higher density ($\rho \sim 0.001\, \mbox{fm}^{-3}$) can be as small as about $1$\,km~\cite{Giunti:2007ry}. One can achieve such extreme high matter densities in neutron stars and supernovae cores which has diameters of a few km.


Here, we investigate, for the first time, the MSW effect in dense quark matter with possible conversion of down-quarks to strange quarks through flavor oscillation. Our motivation to explore such a proposal is driven by the recent observation of possibility of strange quark matter~\cite{Annala:2019puf} considering various astrophysical calculations or possible existence of hyperons in the core of neutron star~\cite{Vidana:2018bdi}. Indeed,  
it is revealed that the interior core of the neutron star indicates characteristics of deconfined phase which can be interpreted as evidence of strange quark matter cores~\cite{Lattimer:2004pg} and (or) existence of strange baryons in neutron star~\cite{Schenke:2021nod,Tolos:2021lta}. 

Our focus in this letter is to put forward the MSW mechanism in quark matter. The main theme of the proposal is the oscillation of quark flavor including medium effects and explore the possibility of resonance oscillations of quark flavors in dense quark matter. 

\noindent

We consider quark flavor oscillation for two generations using down (d) and strange $s$ quarks. Within Standard Model (SM) of particle physics, the usual quarks are classified in three generations where the left-handed ones are structured as isospin doublets
\begin{eqnarray}
 \begin{pmatrix}  u \\ d \end{pmatrix}_L\, , 
 \begin{pmatrix}  c \\ s \end{pmatrix}_L\, ,
 \begin{pmatrix}  t \\ b \end{pmatrix}_L\, 
\end{eqnarray}
while right-handed ones as isospin singlet fields. In general, the down-quark mixes with strange and bottom quarks within three generation picture. As we are limiting our discussion to two generations, the Cabibbo proposal~\citep{Cabibbo:1963yz} of mixing among quarks is given as,
$$\begin{pmatrix}  u^\prime \\ d^\prime \end{pmatrix} = 
  \begin{pmatrix}  u \\ d \cos\theta_C + s \sin\theta_C \end{pmatrix}\, . $$
where $\theta_C$ is Cabibbo mixing angle in vacuum. Here prime-indices denote the weak eigenstate and the unprime quantity correspond to the mass eigenstate. 
The basic idea of quark flavor oscillation is that quarks are produced as flavor eigenstates. Since flavor states can not propagate, they are expressed in mass basis using a unitary transformation. 
The vaccum effect can be understood by relating mass eigenstates $(d,s)$ by their weak eigenstates $(d^\prime, s^\prime)$ as 
\begin{eqnarray}
 \begin{pmatrix}
  d^\prime \\ s^\prime
 \end{pmatrix}
  &=& 
  \begin{pmatrix}
  \cos\theta_C & \sin\theta_C \\
  -\sin\theta_C & \cos\theta_C
 \end{pmatrix}
  \begin{pmatrix}
  d \\ s
 \end{pmatrix} 
\end{eqnarray}
where $\theta_C$ is the quark mixing angle (Cabibbo angle) and $d^\prime$, $s^\prime$ are the flavour (weak) eigenstates. 

The evolution of $d$ and $s$ mass eigenstates with masses $m_d$ and $m_s$ is governed by Schrodinger equation given as,
\begin{equation}
	i\frac{\partial}{\partial t}|\psi(t)\rangle=E|\psi(t)\rangle \,, \quad 
		|\psi (t) \rangle =\left[
	\begin{array}{c}
	 d \\
	 s
	\end{array}
	\right]
\end{equation}
Here, $|\psi(t)\rangle$ is denoted for two-component down-type quark state vector as down-quark and strange-quark mass eigenstates. 
In the limit of relativistic energy approximation $E_i \simeq p+m_i^2/2p \approx E + m_i^2/2E $, we have 
\begin{equation}	
	i\frac{\partial }{\partial t}\left[
	\begin{array}{c}
	 d \\
	 s
	\end{array}
	\right]=\left[\left(
	\begin{array}{cc}
	 E & 0 \\
	 0 & E
	\end{array}
	\right)+\frac{1}{2E}\left(
	\begin{array}{cc}
	 m^2_d & 0 \\
	 0 & m^2_s
	\end{array}
	\right)\right]\left[
	\begin{array}{c}
	 d \\
	 s
	\end{array}
	\right]
\label{eq:schrod2}
\end{equation}
The term proportional to identity has no effect to d-s quark oscillation and hence, can be dropped from the analysis. 
The final mass eigen states are related to corresponding flavor eigenstates by another unitary mixing matrix. 
The propagation and time evolution of down-quarks (part of $SU(2)$ doublets in weak eigenstates) can be expressed in terms of weak eigenstates $d^\prime, s^\prime$ involving Cabibbo mixing angle as, 
	\begin{equation}
	i\frac{\partial }{\partial t}\left[
	\begin{array}{c}
	 d^\prime \\
	 s^\prime
	\end{array}
	\right]=H_{\rm vac}\left[
	\begin{array}{c}
	 d^\prime \\
	 s^\prime
	\end{array}
	\right]
	\end{equation}
with vaccum part of effective Hamiltonian as,
 \begin{eqnarray}
\hspace*{-0.5cm} H_{\rm vac} \hspace*{-0.2cm}&=& \hspace*{-0.2cm}\frac{1}{2 E}
  \left[
	\begin{array}{cc}
	 m^2_d \cos^2\theta_C+m^2_s \sin^2\theta_C & -\cos\theta_C \sin\theta_C \Delta m^2_{ds} \\
	 -\cos\theta_C \sin\theta_C \Delta m^2_{ds} & m^2_d \sin^2\theta_C+m^2_s \cos^2\theta_C
	\end{array}
	\right]
  \nonumber \\
  &=&\frac{\Delta m^2_{ds}}{4E}  \begin{pmatrix} 
    -  \cos2\theta_C & \sin2\theta_C \\
      \sin2\theta_C &  \cos2\theta_C \\
 \end{pmatrix} \big]
 \label{eq:H-vac}
 \end{eqnarray}
with $\Delta m^2_{ds} = m^2_s-m^2_d$ is the mass square difference between strange and down quark and is a positive definite quantity. 

\begin{figure}[htb!]
\includegraphics[width=0.422\textwidth]{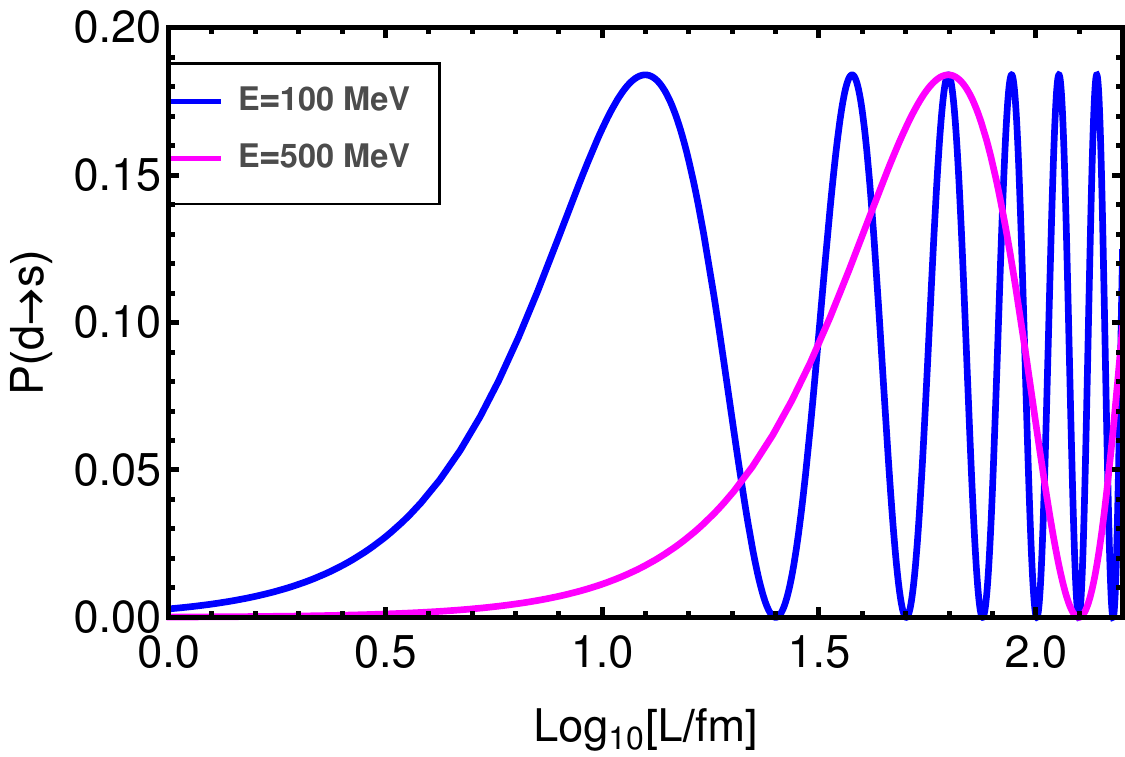}
\caption{Schematic illustration of quark oscillation in vacuum for conversion probability of down quark to strange quark. We consider typical vaccum mixing angle as the Cabibbo angle $\theta_C \simeq 0.22$, energy of down quark flavor is around 100 MeV and distance travelled in terms of fermi. The key point is that a down quark requires to travel a distance of few fermi to convert its flavor to strange quark.}
\label{fig:quark_osciln_vac}
\end{figure}

For relativistic down-type quarks (d,s) with $p \simeq E \gg m^2_{k}$ with $k=d,s$ and $t\approx L$, a typical distance travelled by the quark mass eigenstates, the transition probability for conversion of down-quark flavor to strange-quark flavor is given by~\citep{Sarkar:2008xir}

$$P^{\rm vac}\left(d \xrightarrow{}s\right)=\sin^2{(2\theta_C)}\sin^2{\left(\frac{\Delta m^2_{ds} L}{4E}\right)}$$
The oscillation among quark flavor is possible in vaccum provided two of the following conditions are satisfied: 
\begin{itemize}
\item The mixing angle ($\theta_C$) in vaccum is not be equal to $0$, $n\pi$ or $\frac{n \pi}{2}$. The oscillation amplitude is determined by the Cabibbo mixing angle $\theta_C$. 
\item The mass square difference $\Delta m^2_{ds} =m^2_s-m^2_d \neq 0$. The frequency of the quark flavor oscillation is controlled by this parameter and is large for large value of $\Delta m^2_{ds}$.
\end{itemize}
Let us have an estimate of conversion probability in vaccum. With $\Delta m^2_{ds} \simeq 10^{4}\,\mbox{MeV}^2$, $E\simeq 100$~MeV lead to 
$L_{\rm osc}/2 \simeq 4 \pi$ fermi to have $P^{\rm vac}\left(d \xrightarrow{}s\right)=\sin^2(2\theta_C)\simeq 0.184$. This implies that after travelling a distance of 
$$L_{\rm osc}/2 = \frac{\big(2 \pi \big) E}{\Delta m^2_{ds}}$$ of $4 \pi$ fermi, the probability of finding the quark flavor in the strange quark flavor state $s$ is maximal while after traversing the full length $L_{\rm osc}$, the system is back to its initial state. Thus, e.g., for down quark energy of the order of 100 MeV, probability of the getting it converted to strange quark can be as large as 18 percent after travelling a distance of few tens of fermi even in vaccum which can be understood from the Fig.\ref{fig:quark_osciln_vac}.

Let us consider the medium effects. A quark flavor while passing through the medium of quark matter or hadronic matter, can interact through weak charge current interaction with the medium quarks. This leads to change of mass eigenstates of down and strange quarks resulting eventually in conversion of down quark to strange quarks. The form of charge  current effective Hamiltonian in terms of up and down quarks is given by 
\begin{eqnarray}
 H_{{\rm eff}} &=&\frac{G_F}{\sqrt{2}} \big[\bar{d} \gamma_\mu (1 - \gamma_5) u \big] \big[\bar{u} \gamma^\mu (1-\gamma_5) d \big],
\end{eqnarray}
with $G_F$ as the Fermi constant. Applying Fierz transformation~\citep{Giunti:2007ry} and after  averaging over up-quark background in the non-relativistic limit, i.e., $\langle \overline{u} \gamma^\mu u \rangle \simeq \langle u^\dagger u \rangle \equiv \rho_u$, we get

\begin{eqnarray}
 \bar{H}_{{\rm eff}} &=& \sqrt{2} G_F \rho_u \bar{d}_{L} \gamma^0 d_{L} = 
               V^u_{cc} j^0_d,
\end{eqnarray}
where $d_{L} = \frac{1 - \gamma_5}{2} d$, $j^0_d = \bar{d}_{L} \gamma^0 d_{L}$ and $V^u_{cc}$ is the interaction matter potential given by (non-relativistic limit leading to number density \cite{Pal:1991pm,Kuo:1989qe})
\begin{eqnarray}
 V^u_{cc} = \sqrt{2} G_F \rho_u.
\end{eqnarray}

In the presence of the medium, the evolution equation for the down and strange quarks becomes similar to eq.(5) as,
\begin{eqnarray}
i \frac{\partial}{\partial t}
\begin{pmatrix}
  d^\prime  \\
  s^\prime \\
 \end{pmatrix} 
     =  H_{\rm matt} 
   \begin{pmatrix}
  d^\prime  \\
  s^\prime \\
 \end{pmatrix},
\end{eqnarray}
where, the effective Hamiltonian in medium is given by
 \begin{eqnarray}
  H_{\rm mat} &=& H_{\rm vac}  + \begin{pmatrix}
  V^{u}_{cc}  & 0 \\
  0           & 0
\end{pmatrix}
 \nonumber \\
  &=&
  \begin{pmatrix} 
    - \frac{\Delta m^2_{ds}}{4E} \cos2\theta_C + 2\sqrt{2} G_F \rho_u & \frac{\Delta m^2_{ds}}{4E} \sin2\theta_C \\
     \frac{\Delta m^2_{ds}}{4E} \sin2\theta_C & \frac{\Delta m^2_{ds}}{4E} \cos2\theta_C \\
 \end{pmatrix}. \nonumber 
 \label{matter}
 \end{eqnarray}
\begin{figure}[htb]
\includegraphics[width=0.422\textwidth]{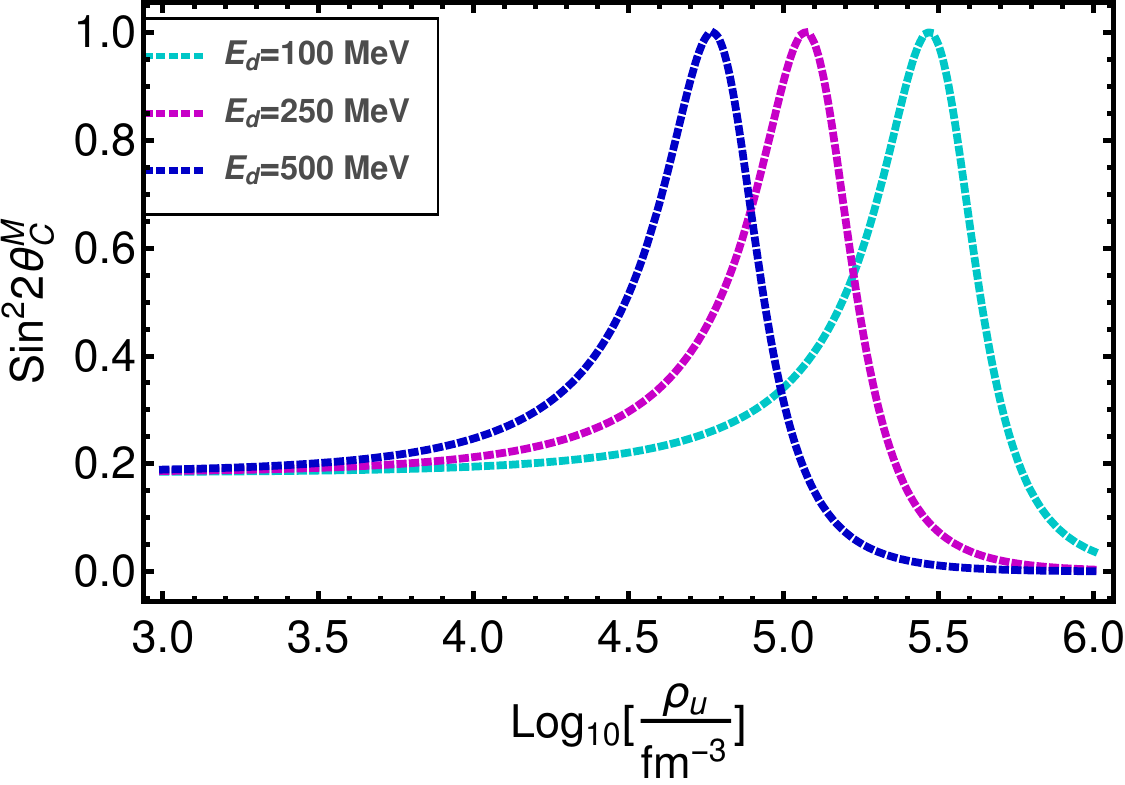}
\caption{Resonant amplification of quark mixing as a result of medium effects with the variation of number density of up-quark $\rho_u$ taken in terms of $\mbox{fm}^{-3}$.}
\label{fig:quark_osciln_resonance}
\end{figure}

Let us introduce a notation characterising the matter effects as,
\begin{eqnarray}
 A = 2\sqrt{2} G_F \rho_u E.
\end{eqnarray}
Using eq.(\ref{matter}), the modified Hamiltonian is read as,
 \begin{eqnarray}
   H_{\rm matt} = \frac{1}{4E}
  \begin{pmatrix} 
      A- \Delta m^2_{ds} \cos2\theta_C  & \Delta m^2_{ds} \sin2\theta_C \\
      \Delta m^2_{ds} \sin2\theta_C &  -A +\Delta m^2_{ds} \cos2\theta_C \\
  \end{pmatrix}. 
  \end{eqnarray}
 The resulting energy eigenvalues of $H_{\rm matt}$ are as follows
 {\small 
\begin{eqnarray}
E^M_{d,s} = \frac{1}{4E}\left[A \mp \sqrt{(-A + \Delta m^2_{ds} \cos2\theta_C)^2 +\bigg(\Delta m^2_{ds} \sin2\theta_C\bigg)^2}\right] 
\end{eqnarray}
}
With the medium effects, the Cabibbo mixing angle $\theta_C$ will be modified as follows,

\begin{eqnarray}
 \tan2\theta^M_C = \frac{\Delta m^2_{ds} \sin2\theta_C}{-A + \Delta m^2_{ds} \cos2\theta_C},
\label{new_mixing}
\end{eqnarray}
\begin{figure}[t!]
\includegraphics[width=0.422\textwidth]{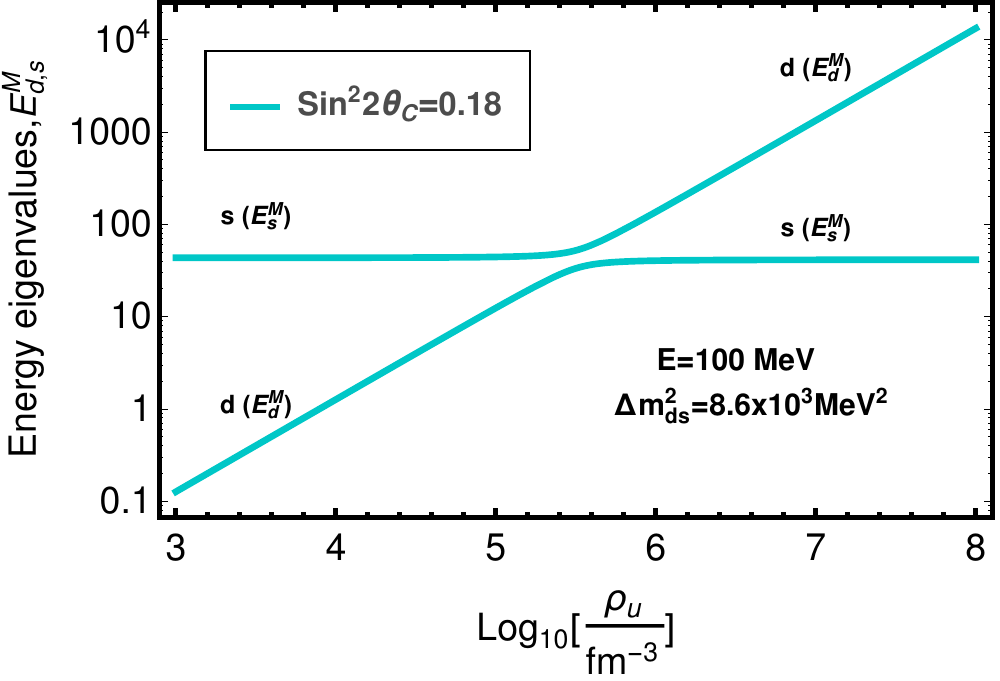}
\caption{Energy eigenvalues for quarks including medium effects and its variation with respect to change of number density of up-quark $\rho_u$.}
\label{fig:quark_osciln_Ens}
\end{figure}

\begin{figure}[h!]
\includegraphics[width=0.422\textwidth]{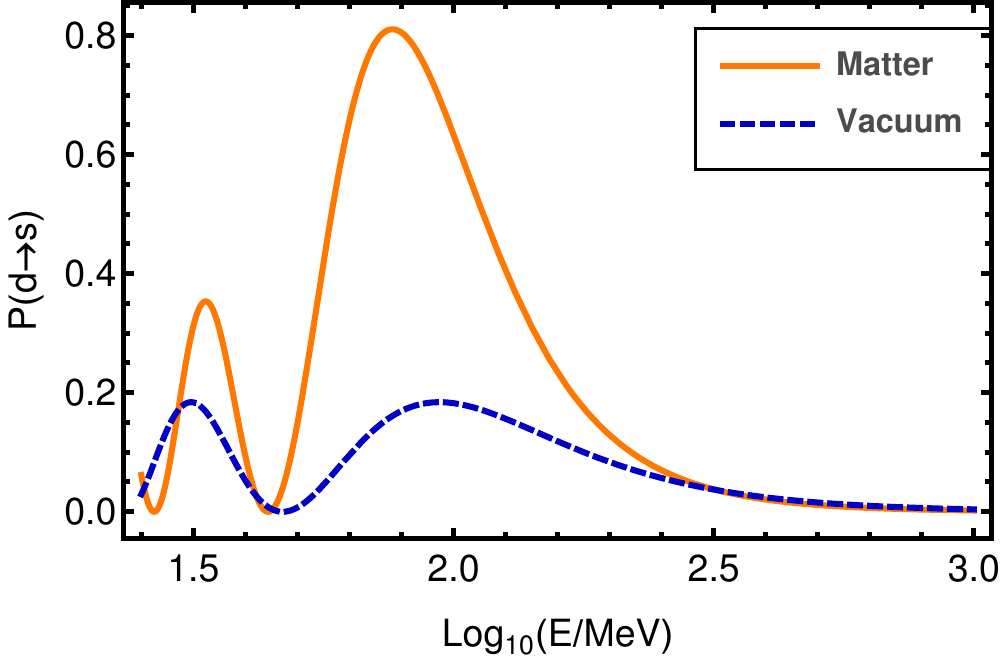}
\caption{Conversion probability of down-quarks converted to strange quark in matter (solid line) and vaccum (dashed line) with variation of energy of quark flavor. The input parameters considered for medium effects are $\Delta m^2_{ds} \simeq 8.6 \times 10^{3}\, \mbox{MeV}^2$ is the mass-square difference between down and strange quark, $\rho_u \simeq 10^{5}\, \mbox{fm}^{-3}$ and distance as $0.1$ fermi. For vaccum, we used the same mass-squared difference parameter but with different distance of 12 fermi for fixing the first oscillation maximum peak at 100 MeV energy of quark flavor.}
\label{fig:prob-vac-mat}
\end{figure}

After all these simplifications, the expression for the probability for $P_{ds}$ becomes
\begin{eqnarray}
P^{\rm mat}\left(d \xrightarrow{}s\right) = \sin^22\theta^M_C \sin^2{\left(\frac{\Delta^M_{ds} L}{4E}\right)}
\end{eqnarray}
Here using the fact that $E_s - E_d \approx (m^2_s - m^2_d)/2E$, we obtain the modified mass squared difference in the presence of matter as
\begin{eqnarray}
\Delta^M_{ds}  = \sqrt{(-A + \Delta m^2_{ds} \cos2\theta_C)^2 +(\Delta m^2_{ds} \sin2\theta_C)^2}.
\label{new_mass}
\end{eqnarray}
It is noted that vaccum oscillation in d-s quarks is not sensitive to sign of $\Delta m^2_{ds}$ and octant of $\theta_C$. However matter effects is sensitive to both of them. 
The resonance condition in presence of medium effects is derived to be
\begin{eqnarray}
 \Delta m^2_{ds} \cos2\theta_C  
  &&=  A \equiv 2\sqrt{2} G_F \rho_u E \nonumber \\
&&\hspace*{-1.2cm}= 2.65 \times 10^{-4} \bigg(\frac{\rho_u}{\mbox{fm}^{-3}}\bigg)\, \bigg(\frac{E}{\mbox{MeV}}\bigg) \, \mbox{MeV}^2  \nonumber \,.
\end{eqnarray}
where $\rho_u$ is the number density of up-quark background with which both down quark is propagating leading to significant medium effects. The  resonance condition for amplification of mixing angle $\sin^2 2 \theta_M$ due to medium effects is displayed in Fig.\ref{fig:quark_osciln_resonance}. The variation of mass eigenstates with eigenvalues $E^{M}_{d,s}$ demonstrating the conversion of down quark to strange quark is presented in Fig.\ref{fig:quark_osciln_Ens}. The conversion probability is presented in Fig.\ref{fig:prob-vac-mat} for illustration of medium effects in quark matter oscillation. 
In the limit $\sin \theta_C \to 0$, the off-diagonal terms can be neglected in comparison to diagonal terms in the effective Hamiltonian in presence of matter. This implies that the resulting energy eigenvalues $E^M_d$, $E^M_s$ and the corresponding eigenstates $(d^\prime, s^\prime)$ are same as their mass eigenstates $(d,s)$. However, due to large medium effects i.e, $\rho_u >> 0$, the eigenvalue of down quark $E^M_d$ can become larger than $E^M_s$ causing conversion of down quark flavor to strange quark flavor. This cross-over occurs at critical density of u-quarks as,
\begin{equation}
	\rho^c_u=\frac{\Delta m^2_{ds} \cos2\theta_C}{2\sqrt{2}G_F  E} ,
	\end{equation}
%
	
\noindent
%
We next examine here the effect of quark flavor oscillation in neutron star. 
The key parameters relevant for $d-s$ quark oscillation using matter effects inside neutron star are
$$\Delta m^2_{ds}, \theta_C, E_d, \rho_u \,.$$
Out of these four parameters, the mass-square difference between strange and down quark ($\Delta m^2_{ds}$) and the Cabibbo mixing angle $\theta_C$ are precisely known while the other two parameters $E_d$ and $\rho_u$ is to estimated for the neutron star medium.  
 
\begin{figure}[t!]
\includegraphics[width=0.422\textwidth]{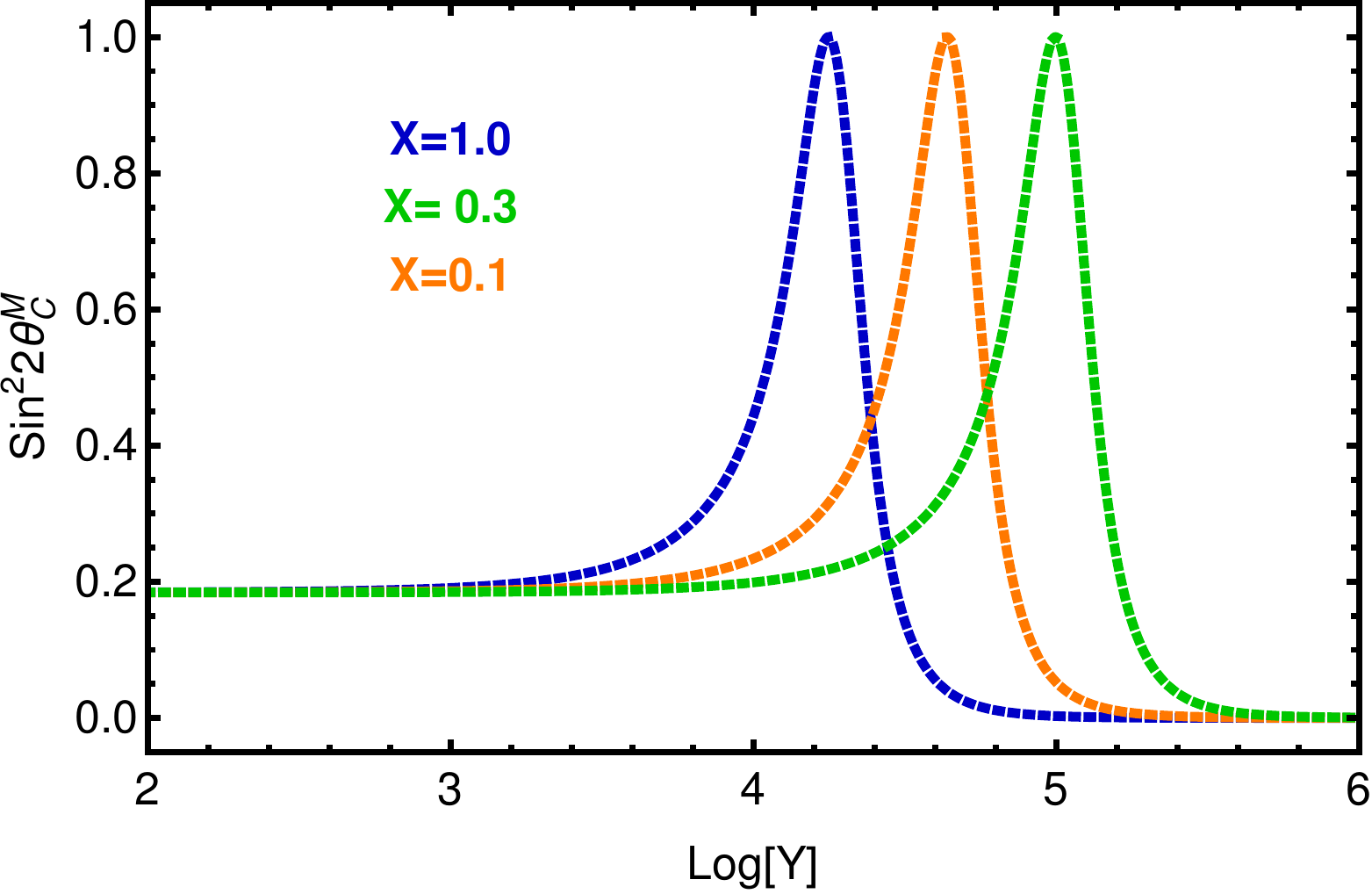}
 \caption{Resonance amplification of mixing angle $\sin^2(2\theta^M_C)$ in presence of high density matter  in terms  $Y$ quantifying up-quark number density in terms of nuclear matter density. Here, we take $X=0.1, 0.3, 1.0$ which is defined as how much down quark can have energy in terms of its fermi momentum.}
\label{fig:resonance_XY}
\end{figure}
The number density of up quarks inside neutron star is written as $\rho_u = Y\, \rho_0$ where $\rho_0 = 0.16 \times \mbox{fm}^{-3}$ is the saturation density of nuclear matter and $Y$ is the parameter defining up-quark density in terms of nuclear matter number density. 
Let us note that up-quark number density $\rho_u \approx \rho$ is the number density at the core of neutron star as there is one up-quark per neutron. We can take $\rho = 5\,\rho_0 = \rho_u$ so that $Y$ is of the order of $5$.
 
The energy of down-quark ($E_d$) can be taken as a fraction of its fermi momentum $k^F_d$ as,
 $$E_d = X\, k^F_d \approx X\, \big(3\,\pi^2\,\rho_d \big)^{1/3} = X\, \big(\frac{3}{2}\,\pi^2\,\rho \big)^{1/3}\,.$$
The modified expression for parameter $A$ is written in terms of $X$ and $Y$ instead of $\rho_u$ and $E_d$ as
\begin{eqnarray}
A &=& \big(2 \sqrt{2}\big) \, G_F\, \big(\frac{3}{2}\,\pi^2\,\big)^{1/3}\, X\,Y^{4/3}\rho^{4/3}_0\,
\end{eqnarray}
The new contribution of matter potential can be inserted in medium effect mixing angle 
$\sin^2(2\theta^M_C)$ and is displayed in Fig.\ref{fig:resonance_XY}.  
For a representative values, $X\simeq 0.5$, $Y\simeq 5$
and $L\simeq 10$~km as the typical radius of the neutron star, the estimated value of conversion probability to have strange quark flavor is $P^M\left(d \xrightarrow{}s\right) \simeq \frac{1}{2} \sin^2{2\theta^M_{C}} \simeq 0.02$. Here we have use the fact that $L$ is much much greater than typical oscillation length, the frequency part of the probability i.e, $\sin^2{\left(\frac{\Delta^M_{ds} L}{4E}\right)} \simeq 1/2$.  
This means that about 2 percent of down quarks are converted to strange quark inside the  neutron star i.e., $\rho_d \approx 0.2 \rho_s$ while travelling a distance from core to the surface of the star. Such a result however depends upon adiabatic approximations i.e, of uniform density throughout the star. Non-adiabatic approximation may change this results. 

%
Let us next compare strange quark number density arising from beta equilibrium condition for the neutron star matter. 
In terms of Fermi momentum and number density, the charge neutrality gives
$$n_{p^+} = n_{e^-}\, \quad \mbox{or,} k_{F,p^+} = k_{F,e^-}\,.  $$

In the other hand, chemical equilibrium yields
$\mu_{n} = \mu_{p^+} + \mu_{e} $. Since the chemical potential is related to the number densities as $\rho_i = \gamma k^3_{F,i}/(6 \pi^2)$ with $k_{F,i} = \sqrt{\mu^2_i-m^2_i}$ with $i=d,s$ and $\gamma$ is the degeneracy factor related to color and spin. Using these basic relations, the number densities of down- and strange quark are related to each other inside neutron star as,
\begin{eqnarray}
 \frac{\rho_d}{\rho_s} = \frac{k^3_{F,d}}{k^3_{F,s}} = \frac{\big( \mu^2_d - m^2_d \big)^{3/2}}{\big( \mu^2_s - m^2_s \big)^{3/2}} \approx 1+ \frac{3}{2}\frac{\Delta m^2_{ds}}{\mu^2_q}
\end{eqnarray}
Using the values of current quark masses, $m_d \simeq 5$~MeV, $m_s \simeq 95$~MeV and $\mu^2_q \simeq \mbox{500\, MeV}^2$, we get 
$$\frac{\rho_d}{\rho_s}=1+0\,.05$$
Thus, the beta equilibrium condition leads to existence of five percent of strange quark number density compare to down-quark number density. 

%
\noindent
%
To summarise we have presented here quark oscillations in vaccum for two generations of quarks similar to neutrino oscillation. It turns out that for a down quark of energy 100~MeV, the oscillation length is a few Fermi's. However, in vaccum such oscillation is not observable as the strong interaction become dominant. On the hand, it is possible to have such flavor oscillation in dense matter. Indeed, following MSW mechanism, it turns out that the resonance oscillation to take place for a very large density of up quark background of the order of $\rho_u \simeq 10^{5}\,\mbox{fm}^{-3}$. 

We finally estimated the conversion probability of down quark to strange quark within typical set of parameters for neutron star and found that about 2 percent of down quark can be converted to strange quark inside the core of the neutron star. This is of the same order as one can expect from the condition of beta equilibrium inside the neutron star. Such an extra possible source of strange quark from flavor oscillation can have interesting consequences regarding equation of state at high densities relevant for neutron star phenomenology. Such an alternate source of strange quarks could possibly lead to higher densities of strange baryons inside neutron star. It can have consequences regarding "Hyperon puzzle" in the context of observation of high mass ($\sim 2\, M_{\odot}$) neutron star~\cite{Vidana:2018bdi}. 
 
%

\bibliographystyle{utcaps_mod}
\bibliography{msw.bib}

\end{document}